\documentclass[12pt]{article}
\usepackage{graphicx}
\begin{document}
{\bf Social Percolation and the Influence of Mass Media}

\bigskip

Ana Proykova$^1$ and Dietrich Stauffer

Institute for Theoretical Physics, Cologne University, 

D-50923 K\"oln, Euroland

\bigskip
\noindent
$^1$ Department of Atomic Physics, University of Sofia, Sofia-1126, Bulgaria

\bigskip

Abstract:
Mass media shift the percolative phase transition observed
in the marketing model of Solomon and Weisbuch.
  

\bigskip
``Social percolation'' \cite{sw99} was invented to describe how a new product
can penetrate a consumer market by word-of-mouth. Imagine a new movie comes out,
and person $i$ goes to the cinema if and only if (s)he hears from neighbours
that the movie quality $q$ is above the own quality expectations $p_i$. 
Initially, $q = 1/2$, and the $p_i$ are distributed randomly between 0 and 1.
If a movie is a success, the next movie has its quality $q$ lowered by 0.001;
if the movie is a flop, the next movie will be better by 0.001. In this way, 
$q$ moves towards the percolation threshold $p_c$ (=0.593 on the square
lattice): self-organized criticality \cite{setal00}.

\begin{figure}[hbt]
\begin{center}
\includegraphics[angle=-90,scale=0.5]{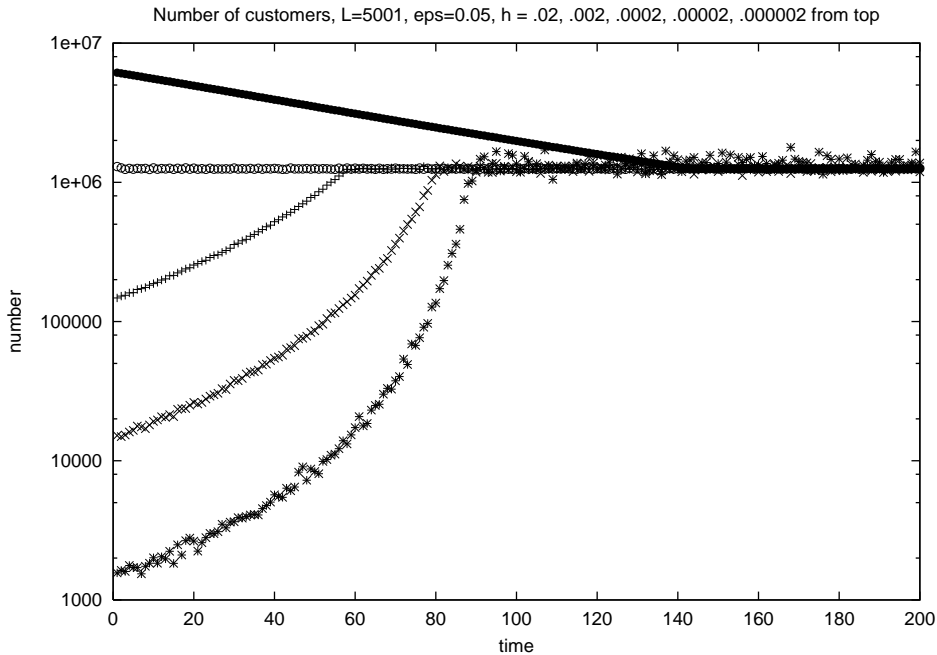}
\end{center}
\caption{
Variation with time of the number of customers for various $h$ = 
0.000002 to 0.02 at fixed $\epsilon = 0.05$, again from 20 lattices
$5001 \times 5001$.
}
\end{figure}

\begin{figure}[hbt]
\begin{center}
\includegraphics[angle=-90,scale=0.5]{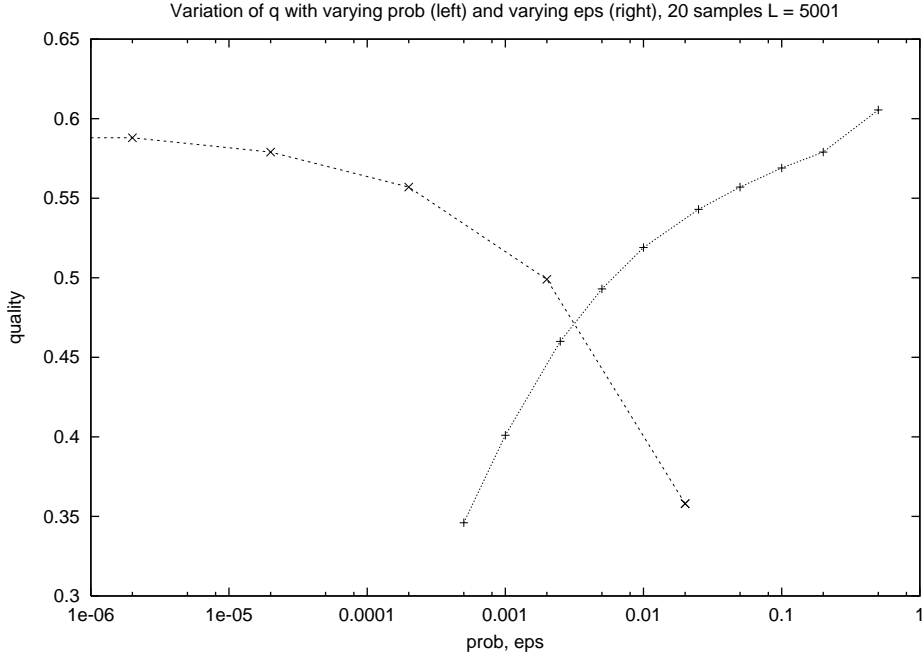}
\end{center}
\caption{
Variation with $h$ for fixed $\epsilon = 0.05$ (left curve) and variation
with $\epsilon$ for fixed $h = 0.0002$ of the final quality $q_f$, averaged
over 20 square lattices $5001 \times 5001$.
}
\end{figure}

\begin{figure}[hbt]
\begin{center}
\includegraphics[angle=-90,scale=0.5]{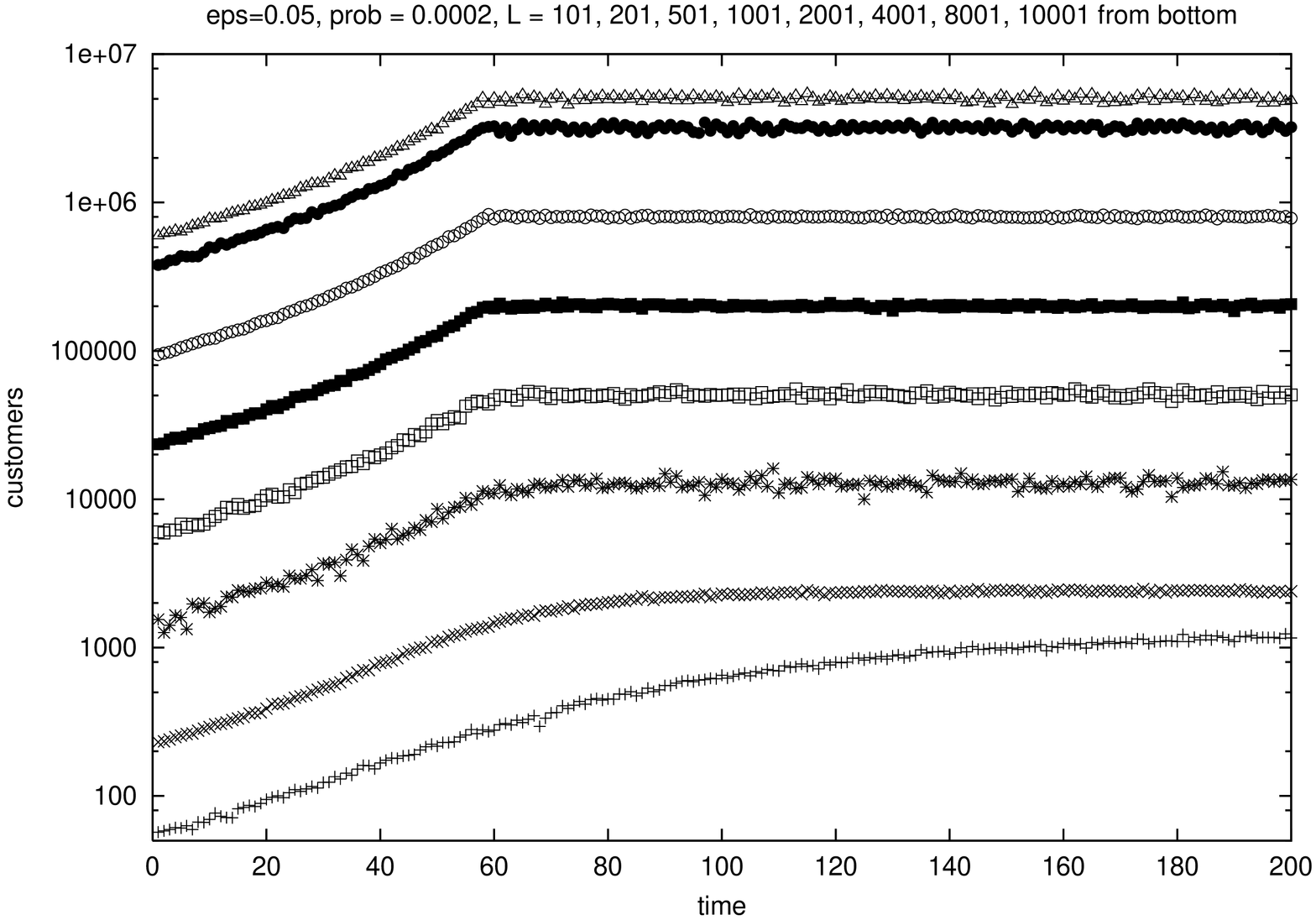}
\end{center}
\caption{
Search for size effects for $101 \times 101$ up to $10001 \times 10001$.
}
\end{figure}

It has been suggested \cite{moukarzel} to include the effect of mass media
through advertizing, in addition to the traditional word-of-mouth propaganda
from the neighbours \cite{sw99,setal00,others}. In percolation theory such
a role is played by the ghost field $h$ which couples each site to a ghost
site with probability $1 - \exp(-h)$, just as telephone links via sattelite
may keep together a widely scattered family. This ghost field $h$ plays the
role of a magnetic field in Ising models and rounds the phase transition at
the critical point. However, since social percolation is a dynamic process,
the analogy of the present simulations with magnetic or ghost fields is not
complete.

We thus assume that with a certain probability $h$ every site $i$ initially
is informed about the quality $q$ of a new movie; the informed sites with own
expectation $p_i$ below $q + 0.001$ visit the movie. Then they tell their 
neighbours, the normal social percolation process starts, the mass
media no longer play any role. (The number 0.001 is added to $q$ to account
for slightly different convincing powers of movie revies compared with neighbour
opinions.) 

When is a movie called a success, causing $q$ to decrease for the next film?
In traditional social percolation without mass media, the criterion was the
existence of a cluster of movie goer extending from one side of the lattice
(Hollywood) to the other (New York). With a ghost field one always has an 
infinite cluster, just as with a magnetic field one gets a finite magnetization.
In our case, through mass media one single site on the border opposite to the 
initially informed border of the lattice can visit the movie. Thus a percolating
cluster no longer means success. Instead, we define a movie to be a success 
if at least a fraction $\epsilon$ of all people go to this movie; typically,
$\epsilon$ is five percent. We no longer have to initialize one lattice border
as being informed; those sites which are convinced by the mass media (typically
$h = 0.0002$) are the starting sites of the word-of-mouth propaganda.

We start with one movie having quality $q$; depending on whether or not the 
threshold fraction $\epsilon$ of movie visits is reached, the quality is changed
to 0.499 or 0.501. Then this movie is put onto the market, with a new 
initial distribution of people convinced my mass media. This process is repeated
again and again for 200 movies, and during this iteration the resulting 
quality $q$ first changes linearly in time until it reaches its stationary
value $q_f$ and fluctuates about this value. To avoid the problems discussed
by Huang \cite{huang}, we do not change the $p_i$ during the simulation of 
one movie; each new movie gets a new distribution of $p_i$.

Fig.1 shows how the number of customers in a market of 25 million varies 
as a function of time (= number of iterations = number of new movies 
introduced to the market). The quality $q$ as a function of time starts 
from zero and moves linearly to its stationary value about which it fluctuates
thereafter.  Fig.2 shows this final quality $q$ as a function of $h$ and 
$\epsilon$. Markets above a million show no significant size effects, Fig.3.
In the stationary state, the distribution of the numbers of customers shows no
clear distinction between hits and flops, in agreement with reality \cite{vany}.

We also coupled the advertising success $h$ with the quality expectations
by assuming a probability $(1+p_i)h$ or $(2-p_i)h$ to believe initially the
mass media, instead of just $h$. However the only role of this coupling, Fig.4,
is the increase of the initial population of buyers
(movie watchers) and the decrease of the time, necessary to convince
reasonable amount of people to go. Most probably this is good for the
trading (movie-making) company.

\begin{figure}[hbt]
\begin{center}
\includegraphics[angle=-90,scale=0.5]{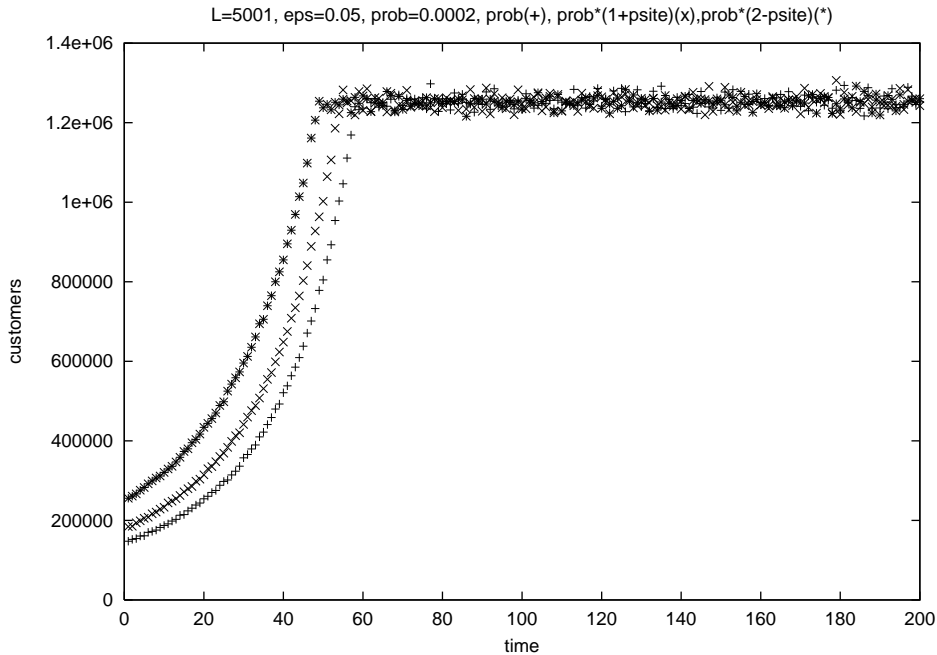}
\end{center}
\caption{
Comparison of homogeneous influence $h$ of advertising (+; as in Figs.1 and 2) 
with an individual advertising success $(1+p_i)h$ (x) and $(2-p_i)h$ (*), 
averaged over 20 lattices
$5001 \times 5001$ at $h = 0.0002, \; \epsilon = 0.05$. 
}
\end{figure}

The plot of $L=2001$ with various $\epsilon$ (not shown) shows the same pattern
already observed in Fig.1 for the decoupled case ($L=5001$). Again the role of
coupling the advertisment to the internal expectations is the 
increase of the initial number of customers.

In summary we see that a moderate amount of advertising decreases the time
after which the number of customers and the quality of the product reaches the 
stationary state (provided the cost of advertising is smaller than the 
resulting profit).

\bigskip 
This work was supported by the Sofia-Cologne university partnership.

\end{document}